\documentclass[a4paper,12pt,twoside]{article} 
\usepackage{a4wide}
\usepackage{graphicx}
\usepackage{rotating}

\begin{document}

{\flushright
CERN-TH/2001-247

FTUV-01-0906

SISSA 46/2001/EP

IFIC/01-53

}

\begin{center}
\noindent{\Large \tt \bf 
%------------------------------------------ Title --------------------

The Earth Mantle-Core Effect in Matter-Induced 
Asymmetries 
for Atmospheric Neutrino Oscillations

%---------------------------------------------------------------------
}\end{center}\vspace{4mm}
\renewcommand{\thefootnote}{\fnsymbol{footnote}}
\noindent{\large
%-------------------------------------- Author(s) --------------------
J. Bernab\'eu$^{a,b}$, S. Palomares-Ruiz$^b$, A. P\'erez$^b$, 
S. T. Petcov$^{c,}$\footnote{Also at: Institute of Nuclear Research and
Nuclear Energy, Bulgarian Academy of Sciences, 1784 Sofia, Bulgaria} 
%---------------------------------------------------------------------
}\vspace{6mm}

\noindent{\small
%------------------------------------ Address(es) --------------------
$^a$  CERN, Geneva, Switzerland
%---------------------------------------------------------------------
\\
%------------------------------------ Address(es) --------------------
$^b$  Departamento de F\'{\i}sica Te\'orica, Universidad de Valencia, 46100 
      Burjassot, Valencia, Spain
%---------------------------------------------------------------------
\\
%------------------------------------ Address(es) --------------------
$^c$ Scuola Internazionale Superiore di Studi Avanzati and
 Istituto Nazionale di Fisica Nucleare,  
I-34014 Trieste, Italy
\\
}

\vspace{6mm}
\renewcommand{\thefootnote}{\arabic{footnote}}
\setcounter{footnote}{0}

\begin{abstract}
Earth medium effects in the three-neutrino oscillations of 
atmospheric neutrinos are observable under appropriate 
conditions. This paper generalizes the study 
of the medium effects and 
the possibility of their observation
in the atmospheric neutrino oscillations
from the case of 
neutrinos traversing  
only the Earth  mantle,
where the density is essentially
constant, 
to the case of atmospheric 
neutrinos crossing  
also the Earth core. 
In the latter case new 
resonance-like effects 
become apparent. We calculate
the CPT-odd asymmetry 
for the survival probability of
muon neutrinos and the observable 
muon-charge asymmetry, taking into account the 
different atmospheric neutrino 
fluxes, and show the dependence 
of these asymmetries on the sign of 
$\Delta m^{2}_{3 1}$ and on the magnitude of 
the mixing angle $\theta_{13}$. A 
magnetized detector with a sufficiently good
neutrino momentum resolution is required
for the observation of the 
muon-charge asymmetry generated by the
Earth mantle-core effect.  
\end{abstract}

\newpage

\vspace{30pt}

  Recently, it was shown \cite{mantle} that
medium effects in the 
three-neutrino oscillations
of atmospheric neutrinos 
crossing the Earth mantle 
become observable under appropriate 
conditions. At the fundamental level, 
their study by means of a magnetized 
detector, able to provide muon charge 
discrimination \cite{Workshop} and energy
resolution, would allow to measure the sign of the 
neutrino mass-squared difference,
$\Delta m^2_{31}$, responsible for
the dominant 
$\nu_{\mu} \leftrightarrow \nu_{\tau}$
and $\bar{\nu}_{\mu} \leftrightarrow \bar{\nu}_{\tau}$
oscillations of the atmospheric  
$\nu_{\mu}$ and $\bar{\nu}_{\mu}$.
Such a study will also 
allow to explore the  
lepton mixing matrix element  $|U_{e 3}| = \sin\theta_{13}$,
which ``connects''
the solar and atmospheric  
neutrino oscillations.
In the limit of 
$\Delta m^2_{21}L/(2E) << 1$,
valid for atmospheric neutrino 
oscillation baselines,
$\Delta m^2_{21} > 0$ 
being the 
neutrino mass-squared difference responsible for 
the oscillations of solar neutrinos,
$L$ the baseline and $E$ the neutrino energy,
the main conclusions  
are based on the following: 
i) the medium effects, which discriminate 
between neutrino 
and antineutrino 
propagation, determine the sign of the 
atmospheric $\Delta m^{2}_{3 1}$ \cite{Barger}; ii) for
$ \sin\theta_{13} \equiv s_{13} = 0$ electron neutrinos decouple 
from the oscillations of the 
atmospheric neutrinos in matter,
whereas they mix with the 
third (heaviest) mass eigenstate 
neutrino and take part in the atmospheric 
neutrino oscillations
if $s_{1 3} \not= 0$, although their mixing with the first 
(lightest) mass eigenstate neutrino still vanishes;
iii) non-resonant medium effects 
are already apparent in the 
sub-dominant 
channels $\nu_e \rightarrow \nu_{\mu}$ and
$\bar{\nu}_e \rightarrow \bar{\nu}_{\mu}$, 
for baselines $L \sim 3000$ km, in both the 
mixing and oscillation phase shift (see also \cite{BeBa00,LBLMSW}); 
iv) in order for the medium effects in 
the muon neutrino survival probability
to be observable, 
the resonant MSW effect in the 
$\nu_{e(\mu)} \rightarrow \nu_{\mu(e)}$ and 
$\bar{\nu}_{e(\mu)} \rightarrow \bar{\nu}_{\mu(e)}$
transitions must be
operational, which requires 
baselines larger than
$L \sim 7000$ km, 
the optimal baseline 
being a function of
the value of $\sin\theta_{1 3}$;
v) taking into account
the initial atmospheric $\nu_{\mu}$, $\bar{\nu}_{\mu}$, and 
$\nu_{e}$, $\bar{\nu}_{e}$
fluxes and the relevant charged current 
neutrino-nucleon deep inelastic scattering
cross-sections, it was shown that
the matter-induced CPT-odd \cite{Lang87}
and CP-odd \cite{BeBa00,LBLMSW,NuFact} asymmetries  
are observable.

   The indicated results were obtained 
for matter of constant density. 
Although there is a 
wide range of Nadir angles (from $33.17^o$ 
to $90^o$), corresponding to 
atmospheric neutrinos crossing the Earth mantle, and
to which the results of the  
study \cite{mantle} apply, 
it is of interest to extend 
the study to the case in which
atmospheric neutrinos cross the Earth core. 
This is the aim of the present 
paper. In our analysis 
we use the two-layer model of the 
Earth density distribution (see, e.g., \cite{Pet}). 
Detailed numerical studies \cite{MP96:2layers} 
(see also \cite{Pet,3nuKP88,MartTommy00})  showed that,
for the calculation of the probabilities 
of interest, the two-layer model of 
the Earth density distribution provides 
a very good (in many cases excellent) 
approximation  
to the more complicated density 
distributions predicted by the existing 
models of the Earth \cite{Earth}.
According to the existing Earth models,
neutrinos which traverse the Earth along 
a trajectory with a Nadir angle 
$\theta_n < 33.17^o$  will cross the 
Earth core. For such trajectories, the distances of 
propagation in the mantle and in the core, $L_m$ 
and $L_c$, are not independent and are given by

\begin{equation}
\label{dist}
\begin{array}{c} \left. \begin{array}{l} 
L_m = R \; \left( \cos{\theta_n} - \sqrt{\frac{r^{2}_{c}}{R^2} - 
\sin^2{\theta_n}} \right) \\[2ex] 
L_c = 2 \; R \; \sqrt{\frac{r^{2}_{c}}{R^2} - \sin^2{\theta_n}}
\end{array} \right\} , \; \sin^2{\theta_n} \leq \frac{r_{c}^{2}}{R^2}
\end{array}
\end{equation}

\noindent
where $R = 6371$ km and $r_c = 3480$ km are the radii of the Earth 
and of the Earth core \cite{Earth}, 
respectively. The total baseline is $L = 2 \, L_m + L_c$, 
with neutrinos propagating in three regions of 
different constant density, mantle-core-mantle, 
the densities of the 
first and third regions being identical.
The probabilities of atmospheric neutrino 
oscillations we will consider are
symmetric with respect to the interchange 
i) of the initial and final
points of the neutrino trajectories, located close to,
or on, the Earth surface, and ii) of the initial and final
flavour neutrinos.
This implies the vanishing of T-odd asymmetries \cite{Ber}.

   The effective neutrino potential differences
in the Earth mantle and in the core, which  
in the case of the 
$\nu_e \rightarrow \nu_{\mu,\tau}$,
and $\nu_\mu \rightarrow \nu_{e}$ 
oscillations of the atmospheric 
neutrinos of interest 
have the well-known form \cite{LW78,Barger80,Lang83}, 

\begin{equation}
V_{m(c)} = \sqrt{2}G_F N^{e}_{m(c)},
\end{equation}

\noindent 
$N^{e}_{m(c)}$ being the electron 
number density in the mantle 
(core)\footnote{In the two-layer model of the Earth density 
distribution, $N^{e}_{m}$ and $N^{e}_{c}$ 
are \cite{Pet,MP96:2layers} the mean 
electron number densities along the 
neutrino trajectory in the mantle and in the core, 
respectively. For the Earth center-crossing neutrinos,
for instance, one has \cite{Earth,Art2}:
$N^{e}_{m} \cong 2.2~{\rm cm^{-3}~N_{A}}$ 
and $N^{e}_{c}\cong 5.4~{\rm cm^{-3}~N_{A}}$,
where ${\rm N_A}$ is the Avogadro number.},  
lead to interesting new 
effects of resonance-like enhancement
of the $\nu_e \rightarrow \nu_{\mu,\tau}$
and $\nu_\mu \rightarrow \nu_{e}$
transitions
beyond the MSW resonance of the mixing
\cite{LW78,Barger80,MS85}. These effects
have been well discussed in \cite{Pet,ChPet,Akh}. 
In the limit $\Delta_{21} \equiv 
\Delta m^{2}_{21} \; L/(2 \; E) = 0$, 
the traceless $2\times 2$ matrix describing
the effective Hamiltonian  
of the two neutrino states
whose evolutions are coupled 
(see, e.g., \cite{3nuto2nu,eff}),
$\nu_e$ and $\nu'$,

\begin{equation}
\label{nuprime}
\nu' = \nu_{\mu} \sin\theta_{23} + \nu_{\tau}\cos\theta_{23}~, 
\end{equation}

\noindent  
where $\theta_{23}$ 
is the mixing angle, which in the limit 
of $\theta_{13} = 0$ controls the 
atmospheric $\nu_\mu \rightarrow \nu_{\tau}$ and
$\bar{\nu}_{\mu} \rightarrow \bar{\nu}_{\tau}$
oscillations, leads to an 
$2\times 2$ unitary evolution matrix of the form
(see, e.g., \cite{JJSaku85,PCWKim93})

\begin{equation}
\label{ev}
U = U_m \; U_c \; U_m, 
\end{equation}

\noindent 
where $U_{m(c)}$ is the evolution operator in the 
Earth mantle (core),

\begin{equation}
\label{evmc}
U_{m(c)} = 
e^{- i(\vec{\sigma} \, \hat{n}_{m(c)}) \phi_{m(c)}} = 
\cos\phi_{m(c)} -i(\vec{\sigma} \, \hat{n}_{m(c)})\sin\phi_{m(c)}. 
\end{equation}

\noindent 
Here 

\begin{equation}
\label{phasemc}
\phi_{m(c)} = \frac{1}{2}\Delta E_{m(c)}~L_{m(c)}
\end{equation}

\noindent 
is the difference between the phases of the
two neutrino energy eigenstates, 
acquired after neutrinos have crossed
the Earth mantle (core), $\Delta E_{m(c)}$ being the 
difference between the
energies of the two states in the mantle (core),

\begin{equation}
\label{DEmc}
\Delta E_{m(c)} = {\Delta m^2_{31} \over 2E} 
\sqrt{\left(\cos2\theta_{13} -
{{2EV_{m(c)}} \over \Delta m^2_{31}} \right)^2 + \sin^2 2\theta_{13}},
\end{equation}

\noindent 
and $\hat{n}_{m(c)}$ is a real unit vector \cite{PCWKim93},

\begin{equation}
\label{nmc}
\hat{n}_{m(c)} = (\sin 2\theta^{m(c)},0,-\cos 2\theta^{m(c)})~,~~
\hat{n}^2_{m(c)} = 1,
\end{equation}

\noindent 
where $\theta^{m(c)}$ is the mixing angle in the
mantle (core), which in the limit of zero neutrino effective
potential difference, $V_{m(c)} \rightarrow 0$, coincides with 
$\theta_{13}$,

\begin{equation}
\label{thetamc}
\cos 2\theta^{m(c)} = {1\over \Delta E_{m(c)}}~
((\Delta m^2_{31}/2E) \cos2\theta_{13} - 
          V_{m(c)}). 
\end{equation}

\noindent 
Using eqs. (\ref{ev})-(\ref{evmc}), 
it is not difficult to 
express the evolution operator of interest 
$U$ in terms of the quantities characterizing 
the evolution of the
neutrino states in the mantle and in the core
(see, e.g., \cite{Akh,JJSaku85}):

\begin{equation}
\label{evf}
U = 
e^{- i(\vec{\sigma} \, \hat{n}) \phi} = 
\cos\phi -i(\vec{\sigma} \, \hat{n})\sin\phi, 
\end{equation}

\noindent 
where 

\begin{equation}
\label{phase}
\cos\phi=
\cos(2\phi_m)\cos\phi_c - \cos(2\theta^c - 2\theta^m)
\sin(2\phi_m)\sin\phi_c,
\end{equation}

\begin{equation}
\label{vectorn}
\hat{n}\sin\phi =
\hat{n}_m\left[\sin(2\phi_m)\cos\phi_c - (\hat{n}_m\cdot\hat{n}_c)
(1-\cos(2\phi_m))\sin\phi_c\right] +
\hat{n}_c\sin\phi_c.
\end{equation}

\noindent 
The two-neutrino oscillation probability
$P (\nu_e \rightarrow \nu') 
= P (\nu' \rightarrow \nu_e) \equiv P_2$,
is determined by the elements of the evolution matrix
$U$, $P_2 = |U_{\nu' \nu_e}|^2$, and can be 
written in the form \cite{ChPet}:   

\begin{equation}
\label{P2}  
P_2 = 
\left(n_1\sin\phi\right)^2 + \left(n_2\sin\phi\right)^2 =
1 - \cos^2\phi - \left(n_3\sin\phi\right)^2,
\end{equation}

\noindent 
where $\cos\phi$ is given by eq. (\ref{phase})  
and  $n_3\sin\phi$ is determined
by eq. (\ref{vectorn}):

\begin{equation} 
\label{n3sinph} 
n_3\sin\phi = 
\cos 2\theta^m[\sin\phi_c \cos2\phi_m \cos(2\theta^c
- 2\theta^m)
+ \cos\phi_c\sin2\phi_m]  \\ 
- \sin\phi_c \sin 2\theta^m\sin(2\theta^c - 2\theta^m).
\end{equation}

 The $\nu_e \rightarrow \nu_{\mu}$ and 
$\nu_e \rightarrow \nu_{\tau}$ transition 
probabilities and the $\nu_e$ survival probability
of interest are related to the probability 
$P_2$ (see, e.g., \cite{3nuto2nu,eff}):

\begin{equation}
\label{probabi}
P (\nu_e \rightarrow \nu_{\mu}) = s^{2}_{2 3} \; P_2 
\hspace{0.3cm}, \hspace{0.7cm}
P (\nu_e \rightarrow \nu_{\tau}) = c^{2}_{2 3} \; P_2 
\hspace{0.3cm}, \hspace{0.7cm}
P (\nu_e \rightarrow \nu_e) = 1 - P_2~.
\end{equation}

\noindent 
where $c_{23}=\cos{\theta_{23}}$ and 
$s_{23}=\sin{\theta_{23}}$. 
We also have
$P (\nu_e \rightarrow \nu_{\mu}) = 
P (\nu_\mu \rightarrow \nu_{e})$.
The probabilities of oscillations of antineutrinos
can be obtained from the corresponding probabilities
of oscillations of neutrinos by replacing
$V_{m(c)}$ by $-V_{m(c)}$ in the 
expressions for the energy 
differences $\Delta E_{m(c)}$  
and the mixing angles $\theta^{m(c)}$ 
in the mantle and in the core, 
eqs. (\ref{DEmc}) and (\ref{thetamc}).

  The conditions for the {\it absolute} ($P_2=1$) maxima 
of the $\nu_e \rightarrow \nu_{\mu}$,
$\nu_{\mu} \rightarrow \nu_{e}$
and $\nu_e \rightarrow \nu_{\tau}$
transition probabilities follow
from the expression (\ref{P2}) for the probability 
$P_2$ \cite{ChPet}:

\begin{equation}
\label{max}
\cos\phi = 0 \hspace{0.3cm}, \hspace{0.7cm} n_3\sin\phi = 0,
\end{equation}

\noindent 
with $\cos\phi$ and $n_3\sin\phi$ 
given by eqs. (\ref{phase}) and (\ref{n3sinph}). 

  In the case of constant density, i.e., 
$V_c = V_m$ and, correspondingly, 
$\Delta E_m = \Delta E_c$, $\theta^c = \theta^m$,
and $\phi = \phi_c + 2 \phi_m = \Delta E_m (L_c + 2L_m)/2$,
the two conditions (\ref{max}) correspond to the 
simultaneous requirement \cite{mantle,ChPet} of both 
a maximum in the oscillating factor
$\sin^2\phi$, namely, $\cos{\phi} = 0$,
and  MSW-resonance in the 
amplitude of the oscillations
$\sin^22\theta^m$,
$n_3 \sin\phi = \cos2\theta^m\sin\phi 
= \pm \cos2\theta^m = 0$. 
There is a crucial difference
between the case of constant density
and the one we are interested in.
When neutrinos cross the mantle, 
the core and the mantle again,
the oscillation phase $\phi$ is no longer linear 
in the baseline L. Instead, it depends 
non-trivially on the 
distances traveled in each layer, $L_m$ and $L_c$.
Similarly, the
condition $n_3\sin\phi = 0$ is a global one, without 
direct correspondence with the MSW-resonance
condition in a given layer. 

  In terms of the oscillation phases 
in each layer, $\phi_m$ and $\phi_c$,
the solution of the conditions (\ref{max}) for the 
absolute maxima of $P_2$ can be written as

\begin{equation}
\label{solu}
\tan{\phi_c} = \frac{1 - \tan^2{\phi_m}}{2 \; (\hat{n}_m \,\hat{n}_c) \; 
\tan{\phi_m}} \hspace{0.3cm}, \hspace{0.7cm}
\tan^2{\phi_m} = \frac{1 - \frac{\Delta E_m}{\Delta E_c} \; 
(\hat{n}_m \,\hat{n}_c)}
{1 + \frac{\Delta E_m}{\Delta E_c} \; (\hat{n}_m \,\hat{n}_c) - 2 \; 
(\hat{n}_m \,\hat{n}_c)^2} ~.
\end{equation}

\noindent
They can be expressed in terms of the 
mixing angles in matter in each of the two layers
(Earth mantle and core) \cite{ChPet}:

\begin{equation}
\label{soluc}
\begin{array}{c}
\tan{\phi_m} = \pm \sqrt{\frac{- \cos{2 \theta^c}}
{\cos{(4 \theta^m -2 \theta^c)}}}~, 
\\[2ex]
\tan{\phi_c} = \pm \frac{\cos{2 \theta^m}}
{\sqrt{- \cos{(2 \theta^c)} \; \cos{(4 \theta^m -2 \theta^c)}}}~,
\end{array}
\end{equation}

\noindent
where the signs in (\ref{soluc}) are correlated. 
Under the assumptions of 
$\cos{2 \, \theta_{1 3}} > 0$ and \linebreak[4]
$\Delta m^{2}_{3 1} > 0$, together with the 
fact that $V_m < V_c$, the domain in  which 
(\ref{soluc}) can be fulfilled is \cite{ChPet}

\begin{equation}
\label{regA}
domain \; A:\; \left\{ \begin{array}{c} 
\cos{2 \theta^c} \le 0 \\[2ex] \cos{(4 \theta^m - 2 \theta^c)} \ge 0
\end{array} \right.
\end{equation}

 The absolute maximum reachable in the domain 
(\ref{regA}) represents a new feature 
of the $\nu_e \rightarrow \nu_{\mu (\tau)}$ and
$\nu_{\mu} \rightarrow \nu_{e}$
transition probabilities
beyond the well-known MSW resonance in the mixing. 
The new enhancement effect disappears
in the limit of constant 
density, when $\theta^c = \theta^m$, 
because the domain ``collapses'' just to the 
resonance condition for the mixing in matter. 
It is not  guaranteed {\it a priori}
that, for a given Nadir angle, corresponding
to neutrino trajectories crossing
the Earth core region, both conditions
(\ref{soluc}) can be satisfied simultaneously,
since the radii of the Earth 
and that of the Earth core and, correspondingly, 
$L_m$ and $L_c$,
and the electron number densities in the
Earth mantle and core, $N^e_{m(c)}$,
are fixed. It was shown in \cite{ChPet} that
solution A is realized 
for the 
$\nu_e \rightarrow \nu_{\mu (\tau)}$ and
$\nu_{\mu} \rightarrow \nu_{e}$
transitions of the Earth core-crossing
atmospheric neutrinos and a rather
complete set of values of
$\Delta m^2_{31}/E$ and 
$\sin^22\theta_{13}$, for which 
both conditions in (\ref{soluc})
hold for neutrino trajectories 
with Nadir angle
$\theta_n = 0^o;~13^o;~23^o;~30^o$, was found.
We will present results for 
the matter-induced 
CP-odd and CPT-odd asymmetries 
as a function of the Nadir angle 
(see below).

  Besides the possible absolute maxima 
described by (\ref{soluc}), with $P_2 = 1$, 
there can be alternative (local) maxima
(in the variables $\phi_m$ and $\phi_c$) 
corresponding to \cite{Pet,ChPet}

\renewcommand{\labelitemi}{i)}
\begin{itemize}
\item
\begin{equation}
\label{mswm}
\cos{2\phi_m} = 0 \hspace{0.3cm}, \hspace{0.7cm} \sin{\phi_c} = 0,
\end{equation}
\begin{center}
in the ``domain'' B: \hspace{1cm} $\cos 2\theta^m = 0$,

with probability \hspace{1cm} $P_2 = \sin^2 2\theta_m = 1$;
\end{center} 
\renewcommand{\labelitemi}{ii)}

\item 
\begin{equation}
\label{mswc}
\sin{\phi_m} = 0 \hspace{0.3cm}, \hspace{0.7cm} \cos{\phi_c} = 0,
\end{equation}
\begin{center}
in the domain C: \hspace{1cm} $\cos{2 \theta^c} \ge 0$,

with probability \hspace{1cm} $P_2 = \sin^2{2 \theta^c}$.
\end{center}
\renewcommand{\labelitemi}{iii)}
\item
\begin{equation}
\label{nolr}
\cos{\phi_m} = 0 \hspace{0.3cm}, \hspace{0.7cm} \cos{\phi_c} = 0,
\end{equation}
\begin{center}
in the domain D: \hspace{1cm} $\cos{(4 \theta^m - 2 \theta^c)} \le 0$,

with probability \hspace{1cm} $P_2 = \sin^2{(4 \theta^m - 2 \theta^c)}$;
\end{center} 
\end{itemize}

\begin{figure}[t]
\begin{center}
\includegraphics[width=10cm]{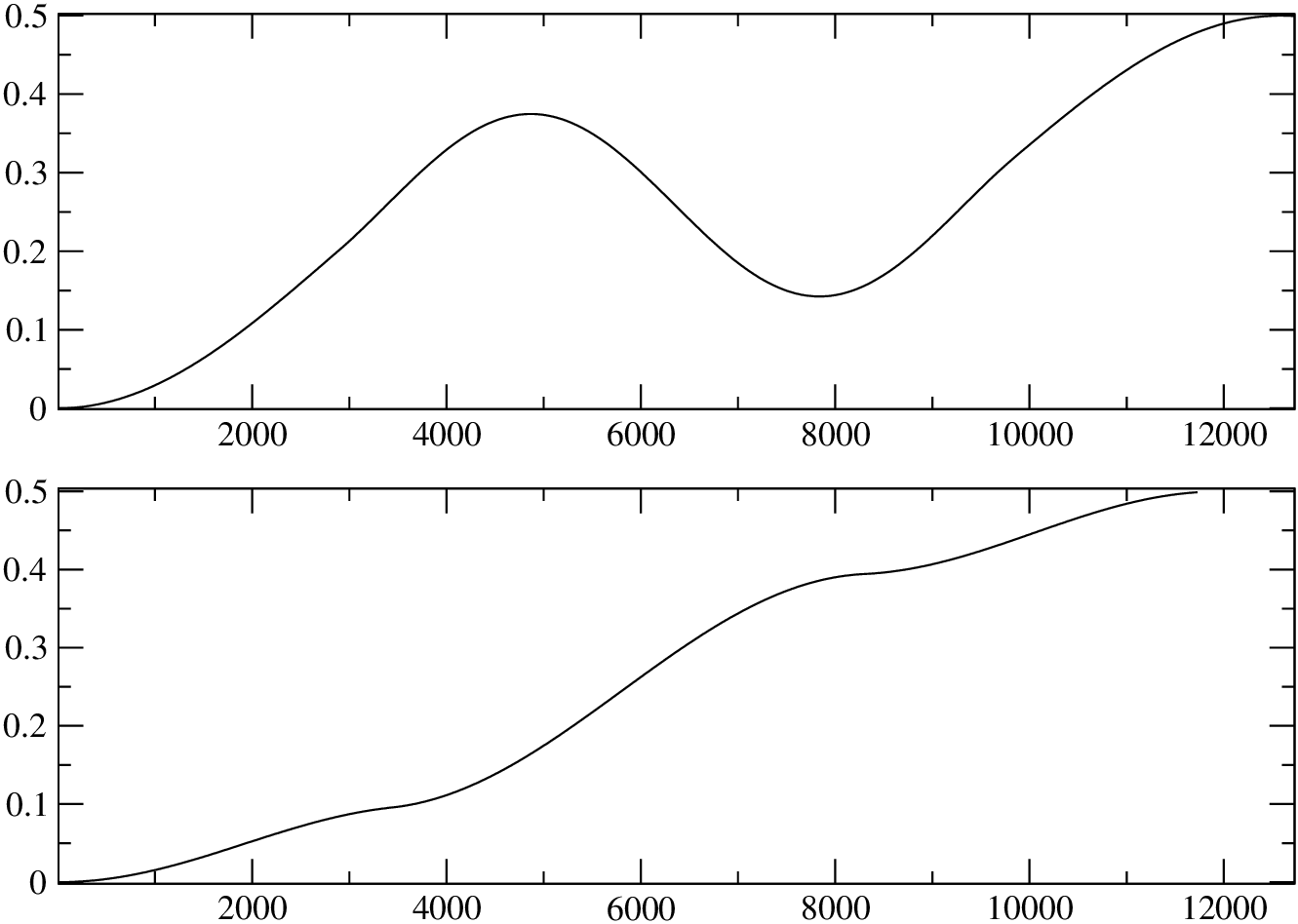}
\caption{\label{evol}Constructive interference 
in the Earth in the channel 
$\nu_e \rightarrow \nu_{\mu}$: 
$P(\nu_e \rightarrow \nu_{\mu})$ as
function of the distance (in km) traveled by the neutrinos
in the Earth, for $\Delta m^{2}_{3 1} = 3.2 \cdot 10^{-3}$ 
eV$^2$ and $\sin^2\theta_{23} = 0.5$. 
The upper panel corresponds to 
solution A for $\theta_n = 0^o$
at $E \simeq 6.6$ GeV and $\sin^2 2\theta_{13}\simeq 0.15$;
the lower panel corresponds to solution B
for $\theta_n = 23^o$ at $E \simeq 5.0$ GeV and
$\sin^2 2\theta_{13} \simeq 0.05$.}
\end{center}
\end{figure}

\noindent 
The maxima of $P_2$ 
described by (\ref{nolr}), known as 
the NOLR solution \cite{Pet}, like 
the absolute maxima of solution A, eq. (\ref{soluc}), 
are due to a 
constructive interference between the 
probability amplitudes of the neutrino 
transitions in the two
different layers - the 
Earth mantle and the core \cite{ChPet}. 
Solution A and solution D
coincide on the border line
$\cos{(4 \theta^m - 2 \theta^c)} = 0$,
which is common to both regions A and D.
The building up of the indicated constructive 
interference is 
illustrated in Fig. \ref{evol}, where we show 
$P(\nu_e \rightarrow \nu_{\mu}) = s_{23}^2 P_2$ 
as a function of the distance traveled by 
the neutrinos in the Earth
in two different cases, 
associated with solutions A and D.
In both cases, we have used \cite{SK} 
$\Delta m^{2}_{3 1} = 3.2 \cdot 10^{-3}$
eV$^2$ and $\sin^2\theta_{23} = 0.5$. 
The two panels correspond to 
solution A for $\theta_n = 0^o$ 
at $\sin^2{2 \theta_{1 3}} \cong 0.15$ and  
$E \cong 6.6$ GeV, and to 
absolute maximum of type B 
(i.e., an absolute maximum
at which $\cos{(4 \theta^m - 
2 \theta^c)} = 0$)
for $\theta_n = 23^o$ at
$\sin^2{2 \theta_{1 3}} \cong 0.05$
and $E \cong 5.0$ GeV \cite{ChPet}.

 Solution $B$ ($C$) defined by (\ref{mswm})  
((\ref{mswc})), 
describes maximum conversion in the mantle (core), 
with no transitions taking place
in the core (mantle). 
The region B is a line 
(in the $E - \sin^22\theta_{13}$ plane)
on which the MSW resonance condition
\cite{Barger80}
is satisfied in the Earth mantle; 
it lies entirely in region A \cite{ChPet}. 
On the border line of 
region $C$, $\cos{2 \theta^c} = 0$,
which is also a border line of region
$A$ and on which solutions $C$ and $A$ 
coincide, the MSW resonance condition 
in the Earth core is fulfilled. 

\begin{figure}[t]
\begin{center}
\includegraphics[width=16cm]{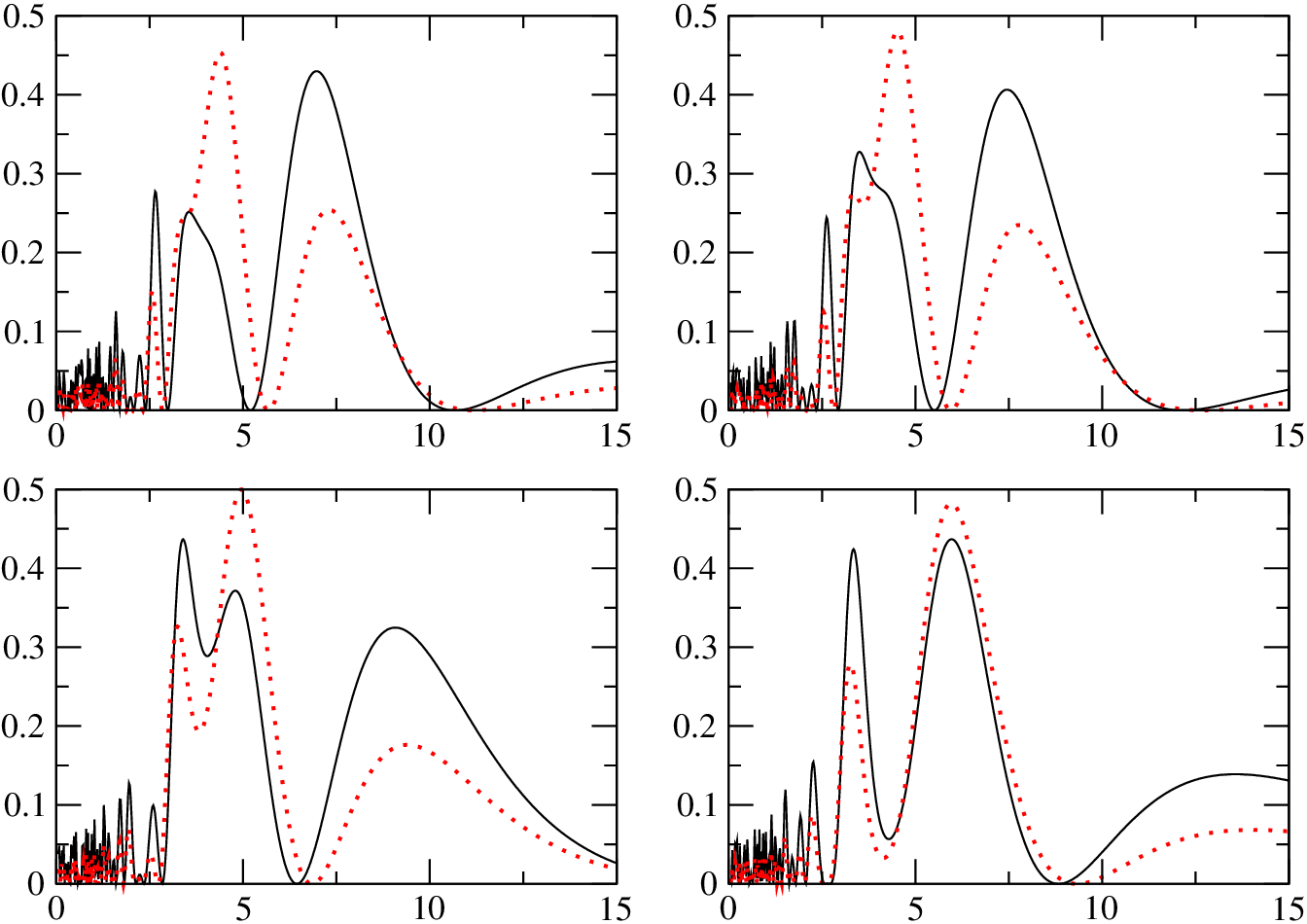}
\caption{\label{pemu} The probability 
$P(\nu_e \rightarrow 
\nu_{\mu})$ for neutrinos
crossing the Earth core, as a function of the 
neutrino energy $E$ 
(in GeV), for    
$\Delta m^{2}_{3 1} = 3.2 \cdot 10^{-3}$ eV$^2$,
$\sin^2 \theta_{23} = 0.5$ and
$\sin^2{2 \theta_{1 3}}$
= 0.10 (solid line), 0.05 (dotted line). 
The upper left and right, and the lower left and right 
panels correspond to neutrino trajectories
with Nadir angle $\theta_n = 0^o;~ 13^o;~23^o;~30^o$,
respectively.}
\end{center}
\end{figure}

  As emphasized in \cite{mantle}, in a situation 
in which the dominant
oscillation channel is 
$\nu_{\mu} \rightarrow \nu_{\tau}$, 
the only method to 
observe the medium effects 
in the $\nu_{\mu}$ survival probability 
(and thus, reach sensitivity to the sign of 
$\Delta m^{2}_{3 1}$ and the value 
of $\theta_{1 3}$) is to operate under 
conditions in which the resonance 
phenomenon in the transitions
$\nu_\mu \rightarrow \nu_{e}$
are observable. For constant 
density in the mantle of the Earth,
this requirement implies a baseline $L \ge 7000$ km, 
i.\ e., a Nadir angle 
$\theta_n \le 56.68^o$. The constant
density approximation
can only be maintained for
$\theta_n \ge 33.17^o$. The question 
is whether for 
$\theta_n < 33.17^o$, one finds, 
for the values of $\Delta m^{2}_{3 1}$
from the atmospheric neutrino 
oscillation region and of
$\theta_{1 3}$ from the region allowed 
by the CHOOZ data, 
solutions to the 
absolute maximum 
conditions (\ref{soluc})
in the energy range relevant 
for the atmospheric neutrinos.
The phenomenon of constructive interference 
between the neutrino transition amplitudes in 
the Earth mantle and in the core was shown 
\cite{Pet,ChPet} to 
take place practically for all Nadir angles, 
associated
with baselines crossing the Earth core,
and for energies of atmospheric neutrinos,
which for $\Delta m^{2}_{3 1} = 3.2\times 10^{-3}~{\rm eV^2}$
lie in the multi-GeV region, $E \cong (3 - 9)$ GeV.
In Figs. \ref{pemu} we show  the 
probability $P(\nu_e \rightarrow \nu_{\mu})$ 
as a function of the neutrino energy $E$
for $\Delta m^{2}_{3 1} = 3.2 \cdot 10^{-3}$ eV$^2$,
$\sin^22\theta_{23} = 0.5$ 
and $\sin^22\theta_{1 3} = 0.10$ (solid line), 
0.05 (dotted line). The four 
panels correspond to neutrino trajectories with Nadir angles 
$\theta_n = 0^o$, $13^o$, $23^o$ and $30^o$,
respectively. 

  The two dominating maxima in the probability 
$P(\nu_e \rightarrow \nu_{\mu})$,
located in the energy intervals $\sim (3 - 5)$ GeV and
$\sim (6 - 8)$ GeV and  clearly seen in Fig. 2,
are for $\theta_n = 0^o;~13^o$,
due to absolute maxima, i.e., solutions of 
eq. (\ref{max}) (or (\ref{soluc})), 
which take place at \cite{ChPet}
$\sin^22\theta^{max1}_{13} = 0.034;~0.039$
and $E^{m-c}_{max1} = 4.4;~4.6$ GeV,
and at $\sin^22\theta^{max2}_{13} = 0.15;~0.17$
and $E^{m-c}_{max2} = 6.6;~7.1$ GeV, respectively.
One of the most important features of the 
results discussed is \cite{Pet} that
the energy of the dominating maxima 
of the probability
$P(\nu_e \rightarrow \nu_{\mu})$,
caused by the mantle-core 
constructive interference effect,
$E^{m-c}_{max}$, lies between 
the energies of the MSW resonance
in the core and in the mantle,
$E^{c}_{res}$ and $E^{m}_{res}$,
which, e.g., 
for $\sin^22\theta_{13} = (0.03 - 0.04)$
read: $E^{c}_{res} \cong 3.9$ GeV 
and $E^{m}_{res} \cong 9.5$ GeV.

 In order to be detected, the medium 
effects discussed
here require a detector with a sufficiently good  
energy resolution. The energies 
at which the dominating maxima of 
$P(\nu_e \rightarrow \nu_{\mu})$ occur
vary somewhat with the Nadir 
angle. Thus, the detector has to 
allow a sufficiently precise 
determination of the direction of the neutrino 
momentum too. It has to be able 
at least to identify with a rather 
good efficiency the events which are due to 
Earth core-crossing neutrinos. 
If these requirements are
fulfilled, the matter effects under discussion
might be measurable.

\begin{figure}[t]
\begin{center}
\includegraphics[width=16cm]{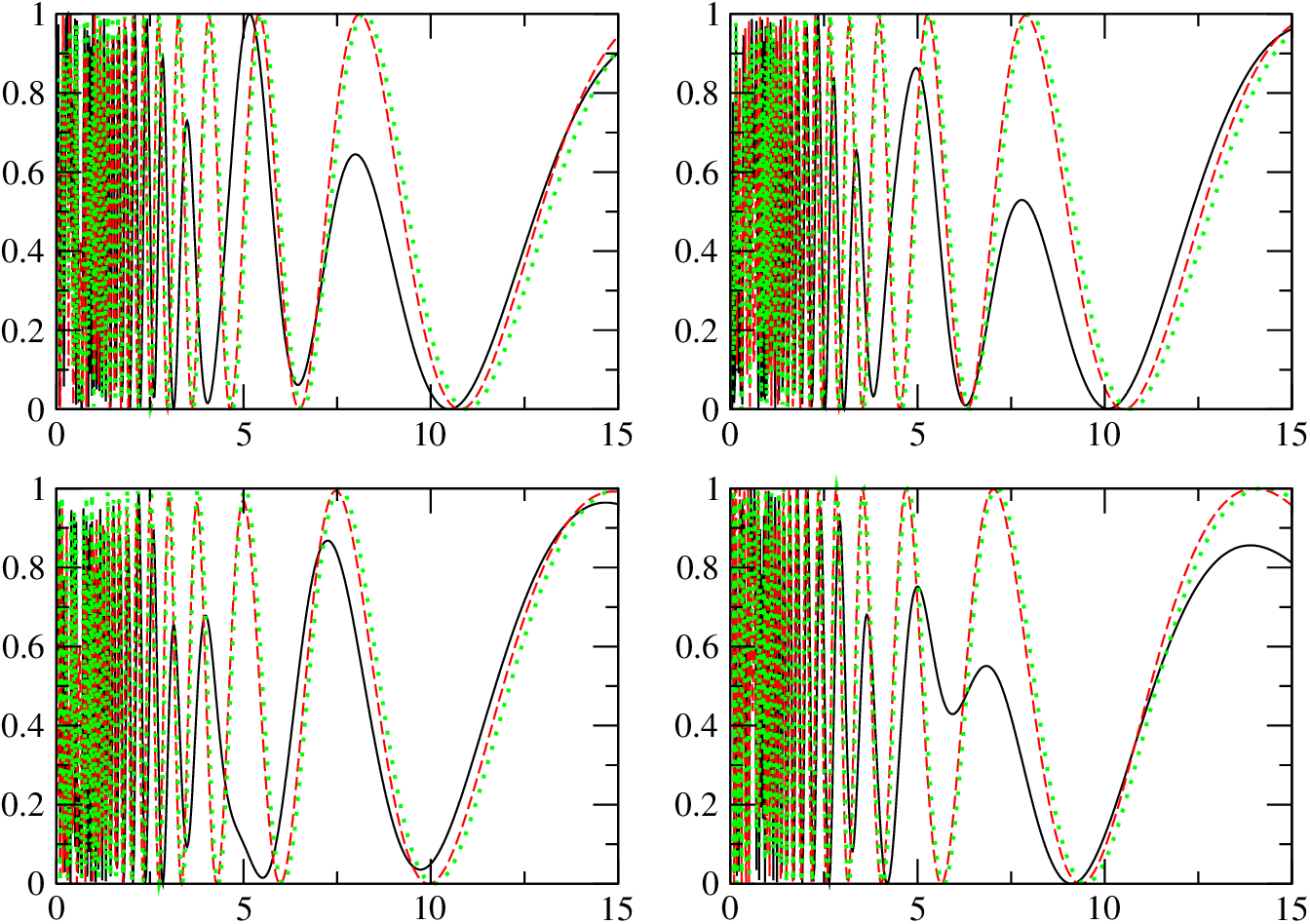}
\caption{\label{pmumu} The survival probabilities, 
$P(\nu_{\mu} \rightarrow \nu_{\mu})$ and 
$P(\bar{\nu}_{\mu} \rightarrow \bar{\nu}_{\mu})$, 
in the case of neutrino oscillations in the Earth 
(solid and dashed lines, respectively) and in vacuum 
(dotted line), as functions of the neutrino 
energy $E$ (in GeV),  
for $\Delta m^{2}_{3 1} = 3.2 \cdot 10^{-3}$ eV$^2$,
$\sin^22 \theta_{23} = 0.5$ and $\sin^2{2 \theta_{1 3}} 
= 0.10$. The upper left, right and the lower left, right panels
correspond to neutrino trajectories with Nadir angle
$\theta_n = 0^o;~13^o;~23^o;~30^o$, respectively.}
\end{center}
\end{figure} 

 Once the Earth mantle-core interference effect
has been studied in the sub-dominant channel 
$\nu_e \rightarrow \nu_{\mu}$, the most 
interesting implication of this  
effect is for the survival probability 
$P(\nu_{\mu} \rightarrow \nu_{\mu})$.
The existence of the absolute maxima 
in $P(\nu_\mu \rightarrow \nu_{e})= 
P(\nu_e \rightarrow \nu_{\mu})$
will lead to an appreciable suppression 
of $P(\nu_{\mu} \rightarrow \nu_{\mu})$
if $\Delta m^{2}_{3 1} > 0$, and 
to no effect in $P(\bar{\nu}_{\mu} 
\rightarrow \bar{\nu}_{\mu})$,
with respect to the 
vacuum case. In Fig. \ref{pmumu} we show 
the dependence of the probabilities 
$P(\nu_{\mu} \rightarrow \nu_{\mu})$ 
and $P(\bar{\nu}_{\mu} 
\rightarrow \bar{\nu}_{\mu})$ on $E$  
for $\Delta m^{2}_{3 1} = 3.2 \cdot 10^{-3}$ eV$^2$, 
$\sin^22 \theta_{23} = 0.5$ 
and $\sin^2{2 \theta_{1 3}}
= 0.10$. The four panels are for Nadir 
angles $\theta_n = 0^o$, $13^o$, 
$23^o$ and $30^o$, respectively. 
The three lines in Figs. \ref{pmumu}  
correspond to $\nu_\mu$ and  
$\bar{\nu}_\mu$ oscillations in the Earth (solid and dashed lines,
respectively) and in vacuum (dotted line).

 To quantify the mantle-core interference effect 
in the $\nu_\mu$ or $\bar{\nu}_\mu$ survival 
probability, we construct the 
matter-induced CPT-odd asymmetry:

\begin{equation}
\label{acpt}
A_{CPT} = \frac{P(\nu_{\mu} \rightarrow \nu_{\mu}) - P(\overline{\nu}_{\mu} 
\rightarrow \overline{\nu}_{\mu})}{P(\nu_{\mu} \rightarrow \nu_{\mu}) + 
P(\overline{\nu}_{\mu} \rightarrow \overline{\nu}_{\mu})}~.
\end{equation}

\begin{figure}[t]
\begin{center}
\includegraphics[width=16cm]{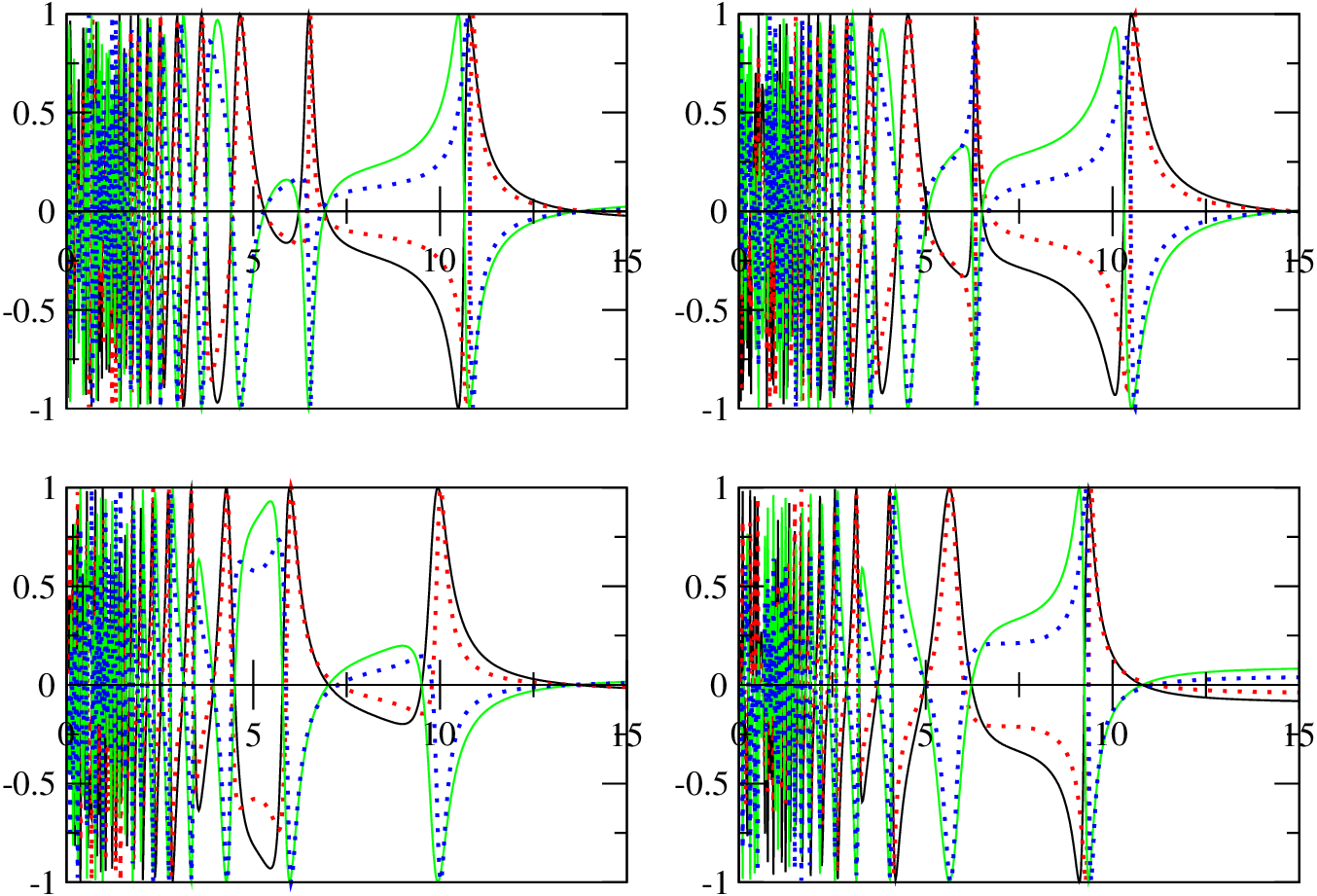}
\caption{\label{amumu} The CPT asymmetry $A_{CPT}$, 
eq. (\ref{acpt}), 
as function of the neutrino energy $E$ (in GeV) for  
$|\Delta m^{2}_{3 1}| = 3.2 \cdot 10^{-3}$ eV$^2$,
$\sin^2 2\theta_{23}$ = 0.5, and 
$\sin^2{2 \theta_{1 3}}$ = 0.1 (solid line), 0.05 (dotted line). 
The solid (doubly thick solid) and dotted 
(doubly thick dotted) lines in each panel correspond to
$\Delta m^{2}_{3 1} > 0$ ($\Delta m^{2}_{3 1} < 0$).
The upper left, right and the lower left, right panels
correspond to neutrino trajectories with Nadir angle
$\theta_n = 0^o;~13^o;~23^o;~30^o$, respectively.}
\vspace{-0.3cm}
\end{center}
\end{figure}

\noindent The CPT-odd probability cancels other \cite{Fogli} 
matter effects common for neutrinos and antineutrinos in
the dominant channel $\nu_{\mu} \rightarrow \nu_{\tau}$. 
The ``theoretical'' asymmetry (\ref{acpt})
is shown in Fig. \ref{amumu} for the same 
values of the Nadir angle used 
in Fig. \ref{pmumu}. In all cases, 
$\Delta m^{2}_{3 1} = 3.2 \cdot 10^{-3}$ eV$^2$, 
$\sin^2 2\theta_{23}$ = 0.5, and 
$\sin^2 2 \theta_{13}$ = 0.10 (solid lines), and 
0.05 (dotted lines).
In the most relevant energy region 
between $\sim$ 4 and $\sim$ 9 GeV, the asymmetry 
is large and with a 
definite sign: negative for 
$\Delta m^{2}_{3 1} > 0$ and positive 
for $\Delta m^{2}_{3 1} < 0$. There is a
change of sign in the interval $E \sim (5 - 6)$ GeV,
whose width depends somewhat on the Nadir angle, 
but the corresponding region is 
rather narrow and the asymmetry in this region 
is relatively small, except for $\theta_n \sim 23^o$. 
This again indicates that to measure the effects
under discussion a sufficient energy 
and Nadir angle resolution 
will be required.

  The {\it observable} muon charge asymmetry in the 
muon events produced by the atmospheric 
neutrinos is a combination of the 
CPT-odd asymmetry (involving the survival probabilities 
$P(\nu_{\mu} \rightarrow \nu_{\mu})$ and
$P(\bar{\nu}_{\mu} \rightarrow \bar{\nu}_{\mu})$) 
and  the CP-odd asymmetry (involving the 
appearance probabilities
$P(\nu_e \rightarrow \nu_{\mu})$ and 
$P(\bar{\nu}_e \rightarrow \bar{\nu}_{\mu})$). 
One can define
 
\begin{equation}
\label{ao}
A = \frac{N(\mu^-;E) - \frac{\sigma_{CC}(\nu_{\mu})}
{\sigma_{CC}(\overline{\nu}_{\mu})} \; N(\mu^+;E)}{N(\mu^-;E) + 
\frac{\sigma_{CC}(\nu_{\mu})}
{\sigma_{CC}(\overline{\nu}_{\mu})} \; N(\mu^+;E)}~,
\end{equation}

\noindent
where $\sigma_{CC}(\nu_{\mu})$ and
$\sigma_{CC}(\overline{\nu}_{\mu})$ are the
relevant charged current neutrino-nucleon 
cross sections, $N(\mu^{\pm};E)$ are the rates of $\mu^{-}$ 
and $\mu^{+}$ events produced by the 
atmospheric neutrinos with energy $E$,

\begin{equation}
\label{nevents}
\begin{array}{c}
N(\mu^-;E) = \sigma_{CC}(\nu_{\mu}) \; \left[ 
\Phi^o (\nu_{\mu};E) \; P(\nu_{\mu} \rightarrow \nu_{\mu}) +
\Phi^o (\nu_e;E) \; P(\nu_e \rightarrow \nu_{\mu}) \right] \\[2ex]
N(\mu^+;E) = \sigma_{CC}(\overline{\nu}_{\mu}) \; \left[ 
\Phi^o (\overline{\nu}_{\mu};E) \; P(\overline{\nu}_{\mu} \rightarrow 
\overline{\nu}_{\mu}) +
\Phi^o (\overline{\nu}_e;E) \; 
P(\overline{\nu}_e \rightarrow \overline{\nu}_{\mu}) \right]
\end{array}
\end{equation}

\noindent
and $\Phi^o(\nu_{l};E)$ and 
$\Phi^o (\overline{\nu}_{l};E)$ are the initial fluxes 
of the atmospheric $\nu_{l}$ and $\overline{\nu}_{l}$, 
$l=e,\mu$, with energy $E$. These fluxes are 
obtained from the code explained in \cite{Naumov}.

\begin{figure}[t]
\begin{center}
\includegraphics[width=16cm]{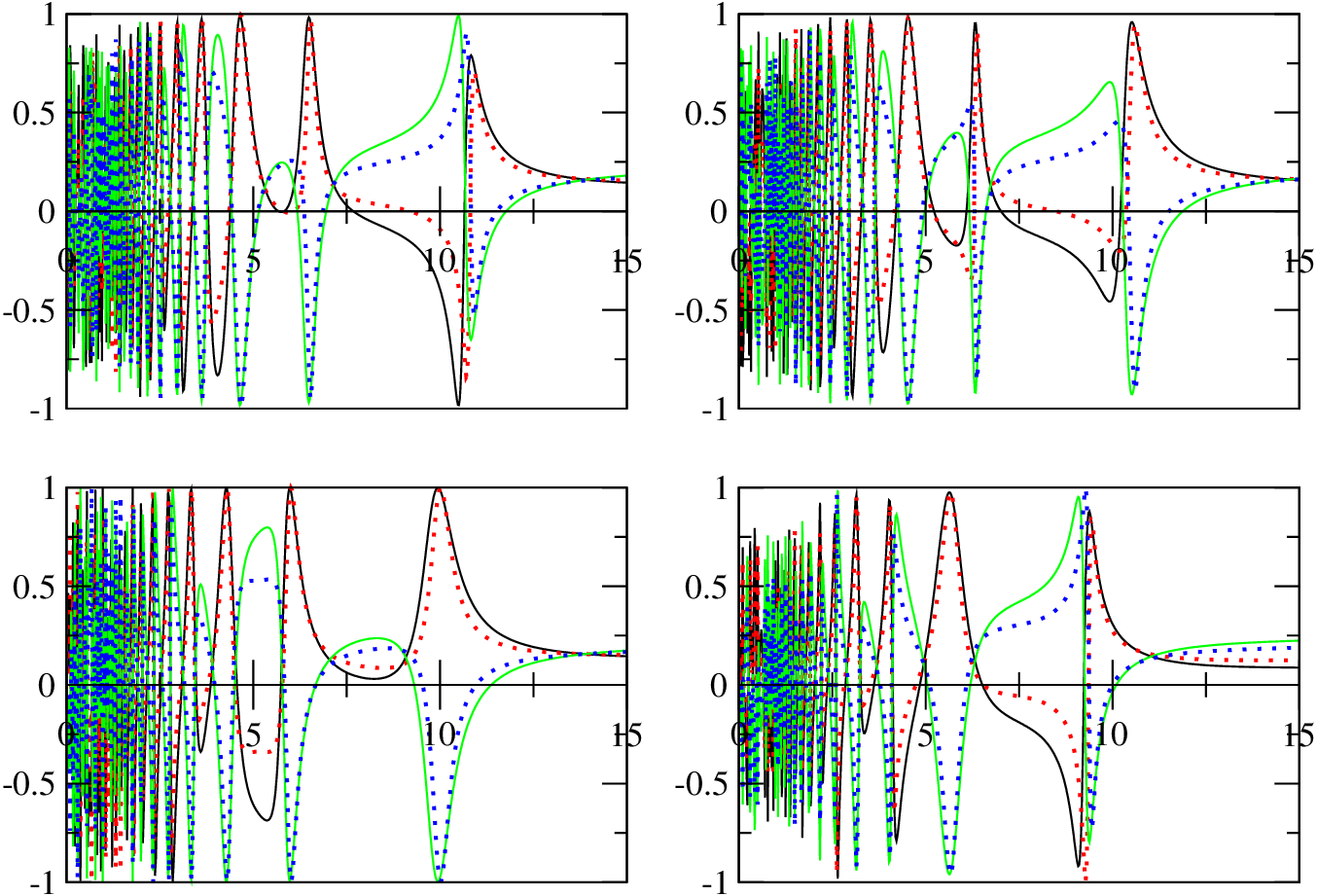}
\caption{\label{gac} The muon charge asymmetry $A$, 
eq. (\ref{ao}),        
as function of the neutrino energy 
(GeV) for  
$|\Delta m^{2}_{3 1}| = 3.2 \cdot 
10^{-3}$ eV$^2$, $\sin^2 2\theta_{23}$ = 0.5,
and $\sin^2{2 \theta_{1 3}}$ = 0.1 
(solid line), 0.05 (dotted line). 
The solid (doubly thick solid) and dotted 
(doubly thick dotted) lines in each panel correspond to
$\Delta m^{2}_{3 1} > 0$ ($\Delta m^{2}_{3 1} < 0$).
The upper left, right and the lower left, right panels
correspond to neutrino trajectories with Nadir angle
$\theta_n = 0^o;~13^o;~23^o;~30^o$, respectively.}
\end{center}
\end{figure}

  The charge asymmetry defined by (\ref{ao}) 
eliminates the asymmetry 
due to the difference of the cross sections
$\sigma_{CC}(\nu_{\mu})$ and
$\sigma_{CC}(\overline{\nu}_{\mu})$ which at 
the energies of interest, and to a good approximation 
\cite{cross-sections}, 
both depend linearly on E. 
Notice that the modulation produced 
by the neutrino fluxes leads 
to a result which is no longer 
symmetric with respect to the horizontal 
axis when changing the sign of 
$\Delta m^{2}_{3 1}$. The net effect, however, is 
the approximate displacement 
of the symmetry axis in the upward direction. The charge asymmetry 
(\ref{ao}) is plotted in Fig. \ref{gac} for the selected Nadir angles  
$\theta_n = 0^o$, $13^o$, $23^o$ and $30^o$.
 The results shown in Fig. \ref{gac} can be compared with those 
in Fig. \ref{amumu} in order to see 
the effects of the contributions 
due to the $\nu_e \rightarrow \nu_{\mu}$
and $\overline{\nu}_e \rightarrow \overline{\nu}_{\mu}$
transitions.  
As is seen in Figs. \ref{amumu} and \ref{gac}, the
main features of the asymmetry $A_{CPT}$, 
discussed earlier,  
are exhibited also by the charge asymmetry A.

  To conclude, the  
propagation of the atmospheric neutrinos
through the mantle and the core 
of the Earth leads to 
new resonance-like effects in the 
$\nu_e (\nu_\mu)\rightarrow \nu_{\mu}(\nu_e)$ 
or $\overline{\nu}_e (\overline{\nu}_\mu) 
\rightarrow \overline{\nu}_{\mu}(\overline{\nu}_e)$
transitions, depending on the sign of 
$\Delta m^2_{31}$. 
This produces observable 
asymmetries between, e.g., 
the atmospheric $\nu_\mu$ and 
$\overline{\nu}_\mu$ 
survival probabilities.
As it was the case of the 
MSW-resonance in the mixing at 
constant density and long enough baselines, 
the measurement of the 
asymmetry between the rates of 
$\mu^{-}$ and $\mu^{+}$ events
due to atmospheric neutrinos
(muon charge asymmetry)
can allow to determine 
the sign of $\Delta m^{2}_{3 1}$.
The muon charge asymmetry 
is rather sensitive to
the magnitude of the ``connecting'' mixing
angle $\theta_{1 3}$ as well, although the wide energy region
around the resonance in the mantle, with stable matter asymmetry,
disappears once the neutrino path crosses the Earth core.
In addition to the muon charge 
discrimination, such measurement requires 
a sufficiently good neutrino 
momentum resolution, both
in magnitude and direction. The implications of these
results on an actual detector are under study \cite{Tab01}.
 
\section*{Acknowledgements}

  We thank F. Dydak, E. Lisi and T. Tabarelli for enlightening 
discussions and V. A. Naumov for providing us 
the code for the calculation of the 
atmospheric neutrino fluxes. J.B. is indebted 
to the CERN Theoretical Physics 
Division for the hospitality extended 
to him during the development of this 
work. A.P. would like to thank the Astronomy and
Astrophysics Department at SUNY (Stony Brook) for 
hospitality.
S. P.-R. is indebted to the Spanish Ministry of 
Education and Culture for a 
fellowship and to SISSA for hospitality. 
This research has been supported by the
Grant AEN-99/0692 of the Ministry of Science 
and Technology, Spain, by Spanish 
DGES Grant PB97-1432, and by the 
Italian MURST 
under the program ``Fisica delle Interazioni
Fondamentali''.

\end{document}